\documentstyle[12pt,psfig,epsf]{article}
\begin{document}
\newcommand{\be}{\begin{equation}}
\newcommand{\ee}{\end{equation}}
\newcommand{\ba}{\begin{eqnarray}}
\newcommand{\ea}{\end{eqnarray}}
\newcommand{\bi}{\bibitem}
\begin{titlepage}
\hbox to \hsize{\large \hfil  IHEP 98-41}
\hbox to \hsize{\large \hfil  June 1998}
\hbox to \hsize{\hfil }
\hbox to \hsize{\hfil }
\hbox to \hsize{\hfil }
\large \bf
\begin{center}
THE MODEL FOR QCD RUNNING COUPLING CONSTANT WITH
DYNAMICALLY GENERATED MASS AND ENHANCEMENT IN THE INFRARED 
REGION\footnotemark 
\footnotetext{
Talk presented at the  Workshop on Methods in Non-Perturbative
Field Theory, 
February 2 -- February 13, 1998, Adelaide, Australia}
\end{center}
\vskip 1cm
\normalsize
\begin{center}
{\bf Aleksey I. Alekseev
}\\
{\small {\it Institute for High Energy Physics,
  Protvino, Moscow Region, 142284 Russia}\\
E-mail: alekseev@mx.ihep.su}
\end{center}
\vskip 1.5cm
\begin{abstract}
Nonperturbative studies of the strong running coupling constant
in the infrared region are discussed. Starting from the analyses
of the Dyson -- Schwinger equations in the gauge sector of QCD,
the conclusion is made on an incomplete fixing of the perturbation 
theory summation ambiguity within "(forced) analytization procedure"
(called also a dispersive approach). A minimal model for 
$\bar\alpha_s(q^2)$ is proposed so that the perturbative time-like
discontinuity  is preserved and nonperturbative terms not only
remove the Landau singularity but also provide the ultraviolet
convergence of the gluon condensate. Within this model,
on the one hand, the gluon zero modes are enhanced (the dual 
superconductor property of the QCD vacuum) and, on the other hand,
dynamical gluon mass generation is realized, with $m_g$ estimated
as $0.6 GeV$. The uncertainty connected with the division into perturbative
and nonperturbative contributions is discussed with the gluon condensate 
taken as an example.
\end{abstract}
\vfill
\end{titlepage}

\section{Introduction}
The infrared region corresponding to large distances is 
of exceptional interest 
by virtue of its responsibility for the confinement mechanism and its
inaccessibility to perturbative methods. A great number of papers
are devoted to the infrared behaviour of gluon Green functions but a 
commonly
accepted opinion is absent in the literature. 

Recently a possibility of power infrared behaviour of the gluon propagator
in the covariant Landau gauge has been discussed,~\cite{Smekal,Atkinson}
with a partial account for ghosts. It was found that in the approximation 
considered, the gluon propagator at small values of Euclidean momenta
vanished (in this case, according to gauge identities, all the 
gluon vertexes
turn out to be singular), the ghost propagator was singular, and a 
running coupling
constant had a large but finite value at zero. 

In the axial gauge in the framework of nonperturbative Baker -- Ball -- 
Zachariasen
(BBZ) approach,~\cite{Baker} which seems to be adequate 
to discuss the possibility of the infrared behaviour  not being too
singular, the problem of consistency of the behaviour
$D(q)\sim(q^2)^{-c}$,
$q^2\rightarrow 0$ was considered.
It has been  shown~\cite{AlekPL,AlekTMF96} that for quite a wide interval
of
non-integer (non-half-integer) values of $c$, $-1<c<3$, there are no 
solutions.
The possibility of "frozen" interaction  in the infrared region
was also considered in the framework of the above approach and 
the result was negative as well.~\cite{AlekArbYF98}

On the other hand, the enhanced  infrared  behaviour
of the gluon propagator, of the form $D(q)\sim 1/(q^2)^2$, $q^2
\rightarrow 0$ is physically motivated, useful in applications 
(e.g. in the quark sector), and it was obtained in
a number of different approaches.~\cite{Pagels,Baker}
This behaviour asymptotically solves the Dyson -- Schwinger equation
in the axial gauge, but with this
we should decline one of the basic assumptions of the BBZ approach
and take into account a transverse part of the triple gluon 
vertex.~\cite{Alek1}
Accordingly, we should decline an appealing iteration scheme~\cite{Baker}
to find solutions of the Dyson -- Schwinger equations for
the higher Green functions.

Versions of the infrared behaviour of the gluon propagator mentioned above
do not exhaust all the possibilities that  can be realized beyond
perturbation theory.
\section{Model for the Running Coupling Constant with 
Nonperturbative Contributions Suppressed in the Ultraviolet Region }
Let us consider the model for $\bar \alpha_s$ introduced 
in Refs.~\cite{AlekArbYF98} and discussed further in Ref.~\cite{22}
We begin with "analytized" expression for the QCD
running coupling constant~\cite{Shir}
\be
\bar\alpha_s^{(1)}(q^2)=\frac{4\pi}{b_0}\left [
\frac{1}{\ln(q^2/\Lambda^2)} +\frac{\Lambda^2}{\Lambda^2-q^2}\right ],
\label{a3}
\ee
obtained with the use of the idea~\cite{Bog} on the cancellation
of the ghost singularity by nonperturbative contributions.
We see that this expression has a nonperturbative tail, with the
behaviour $1/q^2$ at $q^2 \to \infty$. To answer the question whether
this behaviour is admissible,  let us consider
an important physical quantity, namely, the gluon condensate
$K\,=\,{<vac\mid \alpha_s/\pi:F^a_{\mu \nu}\,F^a_{\mu \nu}:\mid 
vac>}\,$. 
Up to the quadratic approximation in the gluon fields,
the gluon condensate
$K$ is defined by nonperturbative contributions in the transverse part of the 
gluon propagator. Normal ordering of  the operators product 
is defined
with respect to perturbative vacuum $|pert>$, as the averaging in 
the expression for the condensate is carried out with true  physical
vacuum state $|vac>$. The following chain of equations describes
the method by which we obtain  closed expression for the gluon condensate:
\begin{equation}
K= \lim_{x\rightarrow y}<vac|\frac{\alpha_s}{\pi}:F^a_{\mu\nu}(x)
F^a_{\mu\nu}(y):|vac>\approx 
\label{4.6}
\end{equation}
\begin{equation}
\approx 2\delta^{ab}\lim_{x\rightarrow y}<vac|\frac{\alpha_s}{\pi}
T[(\partial_\mu A^a_\nu(x)-\partial_\nu A^a_\mu(x))\partial_\mu A^b_\nu(y)]
|vac> - pert =
\label{4.7}
\end{equation}
\begin{equation}
=2\delta^{ab}\frac{\alpha_s}{\pi}\lim_{x\rightarrow y}
(\delta_{\mu\nu}\partial^x_\lambda\partial^y_\lambda - \partial^x_\nu
\partial^y_\mu)
N^{-1}\int dA\, A^a_\mu(x)A^b_\nu(y) e^{S[A]+JA} \mid_{J=0} \, -pert =
\label{4.8}
\end{equation}
\begin{equation}
=2\delta^{ab}\frac{\alpha_s}{\pi}\lim_{x\rightarrow y}
(\delta_{\mu\nu}\partial^x_\lambda\partial^y_\lambda - \partial^x_
\nu\partial^y_\mu)
[D^{ab}_{\mu\nu}(x,y) - D^{pert \,ab}_{\mu\nu}(x,y)]=
\label{4.9}
\end{equation}
\begin{equation}
=2\delta^{ab}\frac{\alpha_s}{\pi}\lim_{x\rightarrow y}
(\delta_{\mu\nu}\partial^x_\lambda\partial^y_\lambda - \partial^x_\nu
\partial^y_\mu)
\int \frac{dk}{(2\pi)^4} e^{-ik(x-y)} [D^{ab}_{\mu\nu}(k)-D^{pert\, ab}
_{\mu\nu}(k)] =
\label{4.10}
\end{equation}
\begin{equation}
=2\delta^{aa}\frac{\alpha_s}{\pi}\int\frac{dk}{(2\pi)^4} (k^2\delta_
{\mu\nu}-k_\mu k_\nu)
[D_{\mu\nu}(k)- D^{pert}_{\mu\nu}(k)]=
\label{4.100}
\end{equation}
\begin{equation}
=\frac{\alpha_s}{\pi^5} \int dk\, (k^2\delta_{\mu\nu}-k_\mu k_\nu )
D^{(0)}_{\mu\nu}(k) [Z(k)-Z^{pert}(k)] =
\label{4.11}
\end{equation}
\begin{equation}
=\frac{3\alpha_s}{\pi^5} \int dk\,[Z(k) -Z^{pert}(k)]=
\label{4.110}
\end{equation}
\begin{equation}
=\frac{3}{\pi^5}\int dk\,[\bar\alpha_s(k^2) -\bar\alpha ^{pert}_s(k^2)]=
\frac{3}{\pi^3}\int dy\,y\bar\alpha^{nonpert}_s(y).
\label{4.12}
\end{equation}

We shall use Eq.~(\ref{4.12}) to evaluate the gluon condensate~(\ref{4.6}).
The transition from Eq.~(\ref{4.100}) to (\ref{4.11}) is made with 
the assumption of proportionality of the nonperturbative part of the 
propagator
to the free one. The projector in Eq.~(\ref{4.11}) removes the terms of 
the free axial gauge propagator $D^{(0)}_{\mu\nu}(k)$ which depend on 
the gauge vector.
Assuming $Z$ is independent of  gauge parameter $y=(k\eta)^2/k^2\eta^2$, 
the employment of the axial gauge lets one 
obtain Eq.~(\ref{4.12}) from~(\ref{4.110}),
thereby expressing the gluon condensate in terms of nonperturbative part
of running coupling constant  $\bar\alpha_s(k^2)$.
The one-loop "analytized" behaviour of Eq.~(\ref{a3}) leads to a
quadratic divergence in Eq.~(\ref{4.12}) 
at infinity
and this is true for
the two- and three-loop expressions~\cite{Shir} of the analytization 
approach.
According to the results of Refs.~\cite{Pagels,Baker,Alek1}
let us add in Eq.~(\ref{a3}) the isolated infrared singular term
of the form $1/q^2$. This term is harmless at zero and it can improve
the behaviour of the integrand at infinity and make the integral
logarithmical divergent. To make the integral~(\ref{4.12}) convergent
at infinity, it is sufficient to add one more isolated singular term
of a pole type with parameters chosen appropriately.
In this sense the model we come to is minimal.
The expression, we obtain for the running coupling constant, is the
following:
\be      
\bar\alpha_s(q^2)\,=\,\frac{4 \pi}{b_0}\Biggl(\frac{1}{
ln(q^2/\Lambda^2)}\,
+\,\frac{\Lambda^2}{\Lambda^2 - q^2}\,+\,\frac{c \Lambda^2}{q^2}\,+
\,\frac{(1-c) \Lambda^2}{q^2 + m_g^2}\Biggr)\,,\label{a45}
\ee
with  mass parameter $m_g=\Lambda/\sqrt{c-1}$.
It is worth  noting that an account of nonperturbative contributions
in Eq.~(\ref{a45}) preserves a perturbative time-like cut of 
Eq.~(\ref{a3}).
With the given value of the QCD scale 
parameter $\Lambda$, the parameter
$c$ can be fixed by the string tension $\kappa$,  
assuming the linear confinement
$V(r)\simeq \kappa r=a^2r$ at $r\to \infty$. 
We define the potential $V(r)$ of static $q\bar q$ interaction
by means of three-dimensional Fourier 
transform of $\bar \alpha_s(\vec q^{\,2})/\vec q{\,^2}$ with 
the contributions of only one dressed gluon exchange taken 
into account. This gives the relation
$
c\Lambda^2=(3b_0/8\pi)a^2
$.
Taking  $a\simeq 0.42\,GeV$, one obtains $c=\Lambda^2_1/\Lambda^2$
where $\Lambda^2_1= 3b_0 \kappa / 8\pi\, \simeq (0.434\, GeV)^2$
($b_0 = 9$ in
the case of 3 light flavours). 
For $m_g$  one obtains 
\be
m_g\,=\,\Lambda^2/\sqrt{\Lambda^2_1-\Lambda^2},\label{s46}
\ee
and the tachion absence condition limits the parameter $\Lambda$ to
$\Lambda < 434 MeV$.

It is seen from Eq.~(\ref{a45}) that the pole singularities are
situated at two points $q^2 = 0$ and $q^2 = - m_g^2$.
This corresponds to the two effective gluon masses, $0$ and $\,m_g$. 
Therefore, the physical meaning of the parameter $m_g$ is not
the constituent gluon mass, but rather the mass of the exited state 
of the gluon. 
\section{Gluon Condensate and Nonperturbative Scale}
The acceptance of the cancellation mechanism for the nonphysical
perturbation theory singularity by the nonperturbative 
contributions leads to the necessity of a supplementary definition of
integral~(\ref{4.12}) near point  $k^2 = \Lambda^2$.
This problem can be reformulated as a problem of dividing
the perturbative and nonperturbative contributions in $\bar\alpha_s$
resulting in the introduction of a parameter $k_0 \approx 1$ GeV.
The following procedure is our definition of the regularized perturbative 
and nonperturbative parts of $\bar\alpha_s(k^2)$:
\be      
\bar\alpha_s(k^2)\,=\,\frac{4 \pi}{b_0}\Biggl(\frac{1}{
\ln(k^2/\Lambda^2)}\,
+\,\frac{\Lambda^2}{\Lambda^2 - k^2}\,+\,\frac{c \Lambda^2}{k^2}\,+
\,\frac{(1-c) \Lambda^2}{k^2 + m_g^2}\Biggr)\,=\label{4.17}
\ee
\be
=\bar\alpha^{pert}_s(k^2)+\bar\alpha^{nonpert}_s(k^2)\,=
\, \bar\alpha^{pert}_{s\,reg}(k^2)+\bar\alpha^{nonpert}_{s\,reg}(k^2),
\label{4.19}
\ee
where
\be      
\bar\alpha^{pert}_{s\,reg}(k^2)\,=\,\frac{4 \pi}{b_0}\Biggl(\frac{1}{
\ln(k^2/\Lambda^2)}\,
+\,\frac{\Lambda^2}{\Lambda^2 - k^2}\theta (k^2_0-k^2)
\Biggr)\,\label{4.20}
\ee
has no power corrections at
$k^2 \rightarrow \infty$ and
\be      
\bar\alpha^{nonpert}_{s\,reg}(k^2)\,=\,\frac{4 \pi}{b_0}\Biggl(
\frac{\Lambda^2}{\Lambda^2 - k^2}\theta (k^2-k^2_0)\,+\,\frac{c 
\Lambda^2}{k^2}\,+
\,\frac{(1-c) \Lambda^2}{k^2 + m_g^2}\Biggr)\,\label{4.21}
\ee
at $k^2 \rightarrow \infty$ has the same power corrections 
as $\bar\alpha^{nonpert}_{s}(k^2)$, namely $\sim \Lambda^6/k^6$.
Thus, we regularize~\footnote{See also Ref.~\cite{Grun} 
where the problem of perturbative 
and nonperturbative contributions to $\bar\alpha_s$ is discussed
and the definition of infrared finite regularized perturbative
part of $\bar \alpha_s$  is suggested.}
the usual perturbation theory in the infrared region.
The arbitrariness of the procedure is parametrized by a scale 
parameter $k_0$.
With account for the stated above, let us calculate the gluon 
condensate~(\ref{4.12}).
We obtain
\be
K(\Lambda,k_0)\,=\,\frac{4}{3 \pi^2}\,\left\{\Lambda^4\,ln\left[
\left(\frac{\Lambda^2_1}{\Lambda^2}-1\right)\left(\frac{k^2_0}{\Lambda^2}
-1\right)\right] +k^2_0\Lambda^2\right\}
\,.\label{4.22}
\ee
In Fig.~1 the dependence of $K^{1/4}$ on $\Lambda$ is shown for different
values of $k_0$ in the interval (0.5 --- 3.0) GeV. For our estimates
we shall use the "standard" value~\cite{ShVZ12} of the gluon condensate
(0.33 GeV)$^4$.
\begin{figure}[t]
\centerline{\psfig{figure=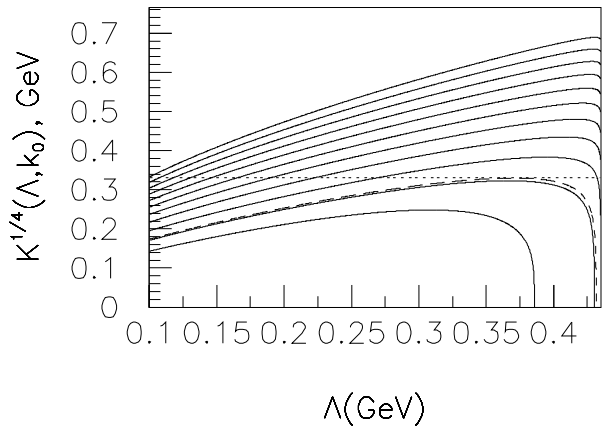,height=8cm,width=10cm}}
\caption{
The dependence of the gluon condensate fourth root $K^{1/4}$ 
on $\Lambda$. The parameter $k_0$ varies from $0.5\,GeV$ (the lowest 
curve) up to $3.0\,GeV$ (the highest one) at an interval of 
$0.25\,GeV$. The dashed line corresponds to the exceptional value 
$k_0 = 0.777\,GeV$. The "standard" level $K^{1/4} = 0.33\,GeV$ is 
indicated.}
\end{figure}
We can see that at  $k_0< \bar k_0 \simeq$ 0.777 GeV gluon 
condensate~(\ref{4.22})
is smaller than its standard value.
It means that there exists a lower limit for the value of parameter $k_0$.
The value $k_0=\bar k_0$ turns out to be exceptional because 
with this choice only one value $\Lambda = \bar \Lambda\simeq$ 375 MeV
provides a necessary value of gluon condensate.
In this case, according to~(\ref{s46}), $m_g = \bar m_g\simeq 0.6$
GeV. At $k_0> \bar k_0$ the two values of $\Lambda$ give the needed value
of the gluon condensate. When $k_0$ increases one of them increases and 
tends to $\Lambda_1$ and the other value of $\Lambda$
decreases and, at  $k_0>2$ GeV, becomes less than 150 MeV.  
\section{Gluon Condensate and Vacuum Energy Density}
Let us represent expression~(\ref{a45}) in an explicitly 
renormalization 
invariant form. It can be done without solving the differential 
renormalization group equations. In this order we write
$\bar\alpha_s(q^2)=\bar g^2(q^2/\mu^2,\; g)/4\pi$ and use the 
normalization condition $\bar g^2(1,g)=g^2$. Then we obtain
the equation for the wanted dependence of the parameter $\Lambda^2$ on
$g^2$ and $\mu^2$:
$$
g^2/4\pi=\frac{4\pi}{b_0}
\left [
\frac{1}{\ln (\mu^2/\Lambda^2)}+\frac{\Lambda^2}{\Lambda^2-\mu^2}+c
\frac{\Lambda^2}{\mu^2}
+\frac{(1-c)\Lambda^2}{\mu^2+m^2_g}
\right ].
$$
For dimensional reasons
$
\Lambda^2=\mu^2 exp\{-\varphi (x)\}$,
where $x=b_0g^2/16\pi^2=b_0\alpha_s/4\pi$, and for the function
$\varphi (x)$ we obtain the equation:
$$
x=\frac{1}{\varphi (x)}+\frac{1}{1-e^{\varphi (x)}}+ce^{-\varphi 
(x)} - \frac{(c-1)^2}{(c-1)e^{\varphi (x)}+1}.
$$
The solution of this equation at $c>1$  is  
function $\varphi (x)$, which has the behaviour $\varphi (x)\simeq 
1/x$ at $x\to 0$ and  $\varphi (x)\simeq -\ln(x/c)$ at
$x\to +\infty$. 
The relation obtained ensures the renormalization invariance of
$\bar\alpha_s(q^2)$. At low  $g^2$, we obtain
$
\Lambda^2=\mu^2\exp\{-4\pi/(b_0\alpha_s)\}$,
which indicates the essentially nonperturbative character of
three last terms in Eq.~(\ref{a45}) and these terms are absent
in the usual perturbation theory.
We will need further the $\beta$-function which can be found by the equation
\be
u\frac{\partial \bar g^2(u,g)}{\partial u} = \beta(\bar g^2).
\label{4.25}
\ee
Here  $u=q^2/\mu^2$. Differentiating the running coupling~(\ref{a45})
in $u$ and assuming $u=1$, we obtain
\be
\beta(g^2)= \frac{16\pi^2}{b_0} \left \{ -x+\frac{1}{\varphi(x)}-
\frac{1}{\varphi^2(x)}+ \frac{1}{(e^{\varphi(x)}-1)^2}-
\frac{(c-1)^2}{((c-1)e^{\varphi(x)}+1)^2}\right \},
\label{4.26}
\ee
which is simplified using equation for $\varphi(x)$.
Then, knowing the behaviour of $\varphi(x)$ at $x \rightarrow 0, \infty$,
we can find
\be
\beta(g^2) \simeq -\frac{b_0}{16\pi^2}g^4+...,\, \, \, g^2\rightarrow 0,
\label{4.27}
\ee
\be
\beta(g^2)\simeq -g^2-\frac{16\pi^2}{ b_0}c(c-2) + O(1/\ln g^2), \, \, \, 
g^2 \rightarrow \infty.
\label{4.28}
\ee
Fig.~2 illustrates the dependence $\beta(g^2)$ for
$c=1,2,...,5$. For all $g^2>0$ the $\beta$-function
is negative definite. Let us consider the trace 
anomaly~\cite{CollinsNielsen}
for the energy - momentum tensor of the gluon field
\be
\Theta_{\mu\mu} =\frac{\beta(g^2)}{2g^2} :F^a_{\mu\nu}F^a_{\mu\nu}:\,.
\label{4.33}
\ee
\begin{figure}[t]
\centerline{\psfig{figure=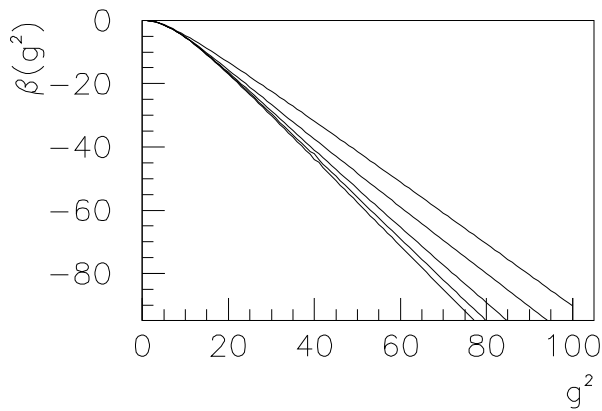,height=8cm,width=10cm}}
\caption{
The $\beta$ -- function for $c=1,2,3,4,5$.}
\end{figure}
According to the definition the vacuum is relativistically invariant. 
So, for the vacuum energy density we have
$
\epsilon_{vac}=(1/4)<vac|\Theta_{\mu\mu}|vac>
$.
For $\beta(g^2)=
-b_0g^4/(16\pi^2)$, $b_0=9$ 
and
$K $= (0.33 GeV)$^4$, the vacuum energy density is
$\epsilon _{vac} \simeq  - (240\,MeV)^4$.
If we introduce quarks, we destroy the nonperturbative vacuum  in some
region (bag).
The vacuum inside the bag is perturbative and its energy density is zero.
The difference of the vacuum energy densities  inside and outside  the bag
is the reason of external pressure on the bag. 

 It is important for us that because of $\beta < 0$, the maximum of gluon
condensate corresponds to the minimum of nonperturbative vacuum energy 
density. For our model $\bar\alpha_s(q^2)$ it means that  the values
\be
\Lambda=\bar\Lambda\simeq 375MeV,\,\,\,
k_0=\bar k_0\simeq 777MeV,\,\,\,
m_g=\bar m_g\simeq 600 MeV
\label{4.36}
\ee
are advantageous from the viewpoint of the energy argument.  
Let us clarify what  the 
dependence of the gluon condensate  on renormalization parameter
$\mu$ is. 
\begin{figure}[t]
\centerline{\psfig{figure=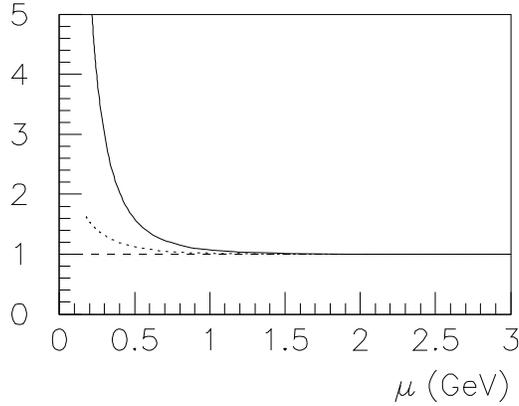,height=8cm,width=10cm}}
\caption{
$K(\mu)/K(\mu\rightarrow \infty)=-(b_0/16\pi^2)g^4/\beta(g^2)$
and $(K(\mu)/K(\mu\rightarrow \infty))^{1/4}$.}
\end{figure}
As a physical quantity, the vacuum energy density  is independent of $\mu$.
The renormalized coupling constant $g$ depends on $\mu$,
$g^2/4\pi=\bar g^2(1,g)/4\pi=\bar\alpha_s(\mu^2)$, so as
$\mu \rightarrow
\infty$ $g^2\rightarrow 0$ and at $\mu \rightarrow 0$ $g^2 
\rightarrow \infty$.
Using~(\ref{4.27}), (\ref{4.28}), we can write
\be
\epsilon_{vac}=\frac{\pi^2\beta(g^2)}{2g^4}<vac|\frac{\alpha_s}{\pi}
:F^a_{\mu\nu}F^a_{\mu\nu}:|vac>=
\label{4.37}
\ee
\be
=-\frac{b_0}{32}<vac|\frac{\alpha_s}{\pi}
:F^a_{\mu\nu}F^a_{\mu\nu}:|vac>\mid_{\mu \rightarrow \infty} =
\label{4.38}
\ee
\be
=-\frac{1}{8}<vac|
:F^a_{\mu\nu}F^a_{\mu\nu}:|vac>\mid _{\mu \rightarrow 0}.
\label{4.39}
\ee
From~(\ref{4.37}), (\ref{4.38}) at $b_0=9$, we have the ratio
\be
\frac{K(\mu)}{K(\mu \rightarrow \infty)}\,=\,-\frac{9}{16\pi^2}
\frac{g^4}{\beta(g^2)},
\label{4.40}
\ee
which describes the dependence of  gluon condensate on $\mu$.
For the $\beta$-function~(\ref{4.26})  in Fig.~3 by a solid line
we show the ratio $K(\mu)/K(\mu\rightarrow \infty)$ as function of
$\mu$. The parameter $c=
1,3476$ corresponds  to the choice 
$\Lambda_1 = 434$ MeV, $\Lambda = \bar\Lambda =375$ MeV.
We can see that for
$\mu >1$ GeV ratio~(\ref{4.40}) is practically unity.
In the same figure the dependence 
$(K(\mu)/K(\mu\rightarrow \infty))^{1/4}$ on  $\mu$ is shown by dots.
\section*{Acknowledgments}
I am grateful to the Special Research Center for the Subatomic 
Structure of Matter
for support and hospitality during my stay at CSSM.
I would like to thank the Organizing Committee of the Workshop
on Methods in Nonperturbative
Field Theory
for all efforts and care.


\begin{thebibliography}{99}
\bibitem{Smekal}
L. von Smekal, A. Hanck, R. Alkofer, {\em Phys. Rev. Lett.} {\bf 79},
3591 (1997);
Report ANL-PHY-8758-TH-97.
\bibitem{Atkinson}
D. Atkinson and J.C.R. Bloch, Report RUGTh-971219; Report RUGTh-980204.
\bibitem{Baker}
     M. Baker, J.S. Ball and F. Zachariasen, {\em Nucl. Phys.} B{\bf 186},
	 531, 560 (1981).
\bibitem{AlekPL}
     A.I. Alekseev, {\em Phys. Lett.} B{\bf 334}, 325 (1995).
\bibitem{AlekTMF96}
     A.I. Alekseev, {\em Teor. Mat. Fiz.} {\bf 106}, 250 (1996).
\bibitem{AlekArbYF98}
A.I. Alekseev, B.A. Arbuzov, {\em Yad. Fiz.} {\bf 61}, 314 (1998);
A.I. Alekseev, B.A. Arbuzov, {\em Mod. Phys. Lett.} A{\bf 13}, 1747 (1998).
\bibitem{Pagels}
     H. Pagels, {\em Phys. Rev.} D{\bf 15}, 2991 (1977); 
     C. Nash and R.L. Stuller, {\em Proc. Roy. Irish Acad.} A{\bf78},
	 (1978); 
     S. Mandelstam, {\em Phys. Rev.} D{\bf 20},3223
     (1979); 
     N. Brown and M.R. Pennington, {\em Phys. Rev.} D{\bf 38},
     2266 (1988);
     D.I. Diakonov and V.Yu. Petrov, {\em Phys. Lett.} B{\bf 242}, 
	 425 (1990);
     K. B\"{u}ttner and M.R. Pennington, {\em Phys. Rev.} D{\bf 52},
     5220(1995);
     A.\,Hauck, L.\,von\,Smekal, R.\,Alkofer, Preprint ANL-PHY-8386-TH,
     1996 (hep-ph/9604430).
\bibitem{Alek1}
     A.I. Alekseev, {\em Yad. Fiz.} {\bf 33}, 516
     (1981); 
     A.I. Alekseev and V.F. Edneral, {\em Yad. Fiz.} {\bf 45},
     1105 (1987).
\bibitem{22}
A.I.  Alekseev,
{\em Talk presented at the XIIth International Workshop on High 
Energy Physics
and Quantum Field Theory, Samara, Russia,
Sept. 4-10, 1997}; Preprint IHEP 97--90, Protvino, 1997;
hep-ph/9802372.
\bibitem{Shir}
     D.V. Shirkov and I.L Solovtsov, {\em Phys.Rev.Lett.} {\bf79},
	 1209 (1997).
\bibitem{Bog}
     N.N. Bogolubov, A.A. Logunov, D.V. Shirkov, {\em Sov. Phys. JETP}
	 {\bf 37}, 805 (1959).
\bibitem{Grun}
     G. Grunberg, hep-ph/9705290.
\bibitem{ShVZ12}
M.A. Shifman, A.I. Vainshtein and V.I. Zakharov,
{\em Nucl. Phys.} B{\bf 147}, 385, 448 (1979).
\bibitem{CollinsNielsen}
J.C. Collins, A. Duncan, S.D. Joglekar,
{\em Phys. Rev.} D{\bf 16}, 438 (1977);
N.K. Nielsen, {\em Nucl. Phys.} B{\bf 120}, 212 (1977). 

\end{thebibliography}
\end{document}